\documentstyle[12pt,aaspp4]{article}

\begin{document}
\title{Low-Mass Binary Induced Outflows from Asymptotic Giant Branch Stars}
\author{J. Nordhaus\altaffilmark{1,2}, E. G. Blackman\altaffilmark{1,2}}
\affil{1. Dept. of Physics and Astronomy, Univ. of Rochester,
    Rochester, NY 14627}\affil{2. Laboratory for Laser Energetics, Univ. of Rochester, Rochester, NY 14623}
\begin{abstract}
A significant fraction of planetary nebulae (PNe) and proto-planetary nebulae (PPNe) exhibit aspherical, axisymmetric structures, many of which are highly collimated.  The origin of these structures is not entirely understood, however recent evidence suggests that many observed PNe harbor binary systems, which may play a role in their shaping.  In an effort to understand how binaries may produce such asymmetries, we study the effect of low-mass ($<0.3$ $M_\odot$) companions (planets, brown dwarfs and low-mass main sequence stars) embedded into the envelope of a $3.0$ $M_\odot$ star during three epochs of its evolution (Red Giant Branch, Asymptotic Giant Branch (AGB), interpulse AGB).  We find that common envelope evolution can lead to three qualitatively different consequences:  (i) direct ejection of envelope material resulting in a predominately equatorial outflow, (ii) spin-up of the envelope resulting in the possibility of powering an explosive dynamo driven jet and (iii) tidal shredding of the companion into a disc which facilitates
a disc driven jet.  We study  how these features depend on the 
secondary's mass and discuss observational consequences. 
\end{abstract}
\section{Introduction}
Deviation from spherical symmetry in planetary nebulae (PNe) and 
protoplanetary nebular (PPNe) 
can be pronounced  (\cite{1997ApJS..112..487S}, Balick \& Frank 2002, Bujarrabal et al. 2004).  Understanding the origin of the asymmetries 
is an ongoing aim of current research.

A variety of scenarios have been proposed to explain the transition from progenitor to planetary nebula.  As low- and intermediate-mass main sequence stars evolve onto the Asymptotic Giant Branch (AGB), enhanced stellar wind mass loss leads to depletion of the hydrogen envelope around the central core.  Recent surveys (\cite{2000ASPC..199..209S}, \cite{2002RMxAC..13..133S}) of AGB and post-AGB stars have revealed spherical symmetry 
leading to the conclusion that any shaping process must occur in a relatively short time just before the birth of the PNe in the PPNe, or  AGB phase (Bujarrabal et al. 2001).

Bipolar outflows in either the AGB or post-AGB phase could produce such structures.  However, the mechanism by which the bipolar winds are produced remains 
to be fully understood.  Binary interactions, large-scale magnetic fields and high rotation rates of isolated AGB stars may all play some role in  explaining the observed structures:  
\cite{1994A&A...282..554F} appealed to a superwind induced by a thermal helium flash as a possible production mechanism for bipolar planetary nebulae.   \cite{2002A&A...386..885S} argued that such a model does not account for the observational link between aspherical mass loss and asymptotic wind.  Such an observational correlation could be explained by a binary system in which the companion is a low-mass star or brown dwarf (\cite{S2004}).  

AGB and post-AGB remnant central stars are known to possess magnetic fields (\cite{2004ASPC..313..186B} and \cite{2005A&A...432..273J}).  In addition, direct evidence of a magnetically collimated jet in an evolved AGB star has been detected (\cite{2006Natur.440...58V}), further 
suggesting some dynamical role for magnetic fields.
Magnetic outflows from single stars have been proposed as mechanisms
for  shaping PPNe and PNe (Pascoli 1993; Blackman et al. 2001a).
However, single star models may be unable to sustain the necessary Poynting flux required to 
maintain an outflow through the lifetime of the AGB phase unless differential
rotation is reseeded by convection 
or supplied by a binary (Blackman 2004, \cite{2006PASP..118..260S}). 
A model in which a disc driven magnetic dynamo driven outflow 
is sustained by accretion from a shredded secondary
was pursued  in (Blackman et al. 2001b).

In this respect, it is noteworthy that recent studies support the claim that most, if not all, observed planetary nebulae are the result of a binary interaction (\cite{2004ApJ...602L..93D}, \cite{2004ASPC..313..515S}, \cite{DM}, \cite{Mauron}).  While this conclusion is based on observations and population synthesis studies, \cite{Soker2006} points out that the corresponding formation rate of PNe from such studies is 1/3 the formation rate of white dwarfs.  This supports the claim that binary stars may produce the more prominently observed planetary nebulae (\cite{2005AJ....130.2717S}).  

The question of just how  a binary shapes a 
PNe or PPNe remains a topic of active research.  
In this paper, we explore the effects of an embedded low-mass companion inside the envelope of a 3 $M_\odot$ star during three epochs of its evolution off the main sequence (Red Giant Branch (RGB), AGB and interpulse AGB).  
A common envelope (CE) facilitates mass ejection in several ways:  The in-spiral of the secondary toward the core deposits orbital energy and angular momentum in the envelope.  This directly ejects and/or spins up  the envelope.  
In the latter case, any enhanced differential rotation could aid 
in  magnetic field generation, which in turn could  drive mass loss.  
In section 2, we describe the stellar models used and derive the basic
equations for in-spiral and for the transfer of energy and angular momenutum
from the secondary to the envelope.
Results for different evolutionary epochs are discussed in section 3. We present observational implications and applications to specific systems in section 4
and conclude in section 5.

\section{Common Envelope Evolution}
Under certain conditions, Roche lobe overflow in close binary systems results in both companions immersed in a CE (\cite{1976IAUS...73...75P}, \cite{1993PASP..105.1373I}).  Once inside, velocity differences between companion and envelope generate a drag force that acts to reduce the orbital separation of the companion and core.  Orbital energy is deposited into the envelope during the in-spiral process. Some of this energy is radiated away while
the rest is available to reduce the gravitational binding energy of the envelope.  The efficiency with which orbital energy unbinds envelope matter  is of central importance to CE evolution.  This is commonly incorporated into  a parameter, $\alpha$, which represents the fraction of orbital energy available for mass ejection as follows: 
\begin{equation}
E_{bind}=\alpha{\Delta{E_{orb}}},
\end{equation}
where $\Delta{E_{orb}}$ is the change in orbital energy of the binary and $E_{bind}$ is the energy required to unbind envelope material.  In principle, knowledge of the binding energy and $\alpha$ determines how much material is ejected, and in the case of complete envelope ejection, final binary separation distances.  Such studies have been performed for a variety of different systems under a range of conditions of which the following are a small sample (\cite{1995ApJ...447..656Y}, \cite{2000A&A...360.1043D}, \cite{2000ARA&A..38..113T}, \cite{2004ApJ...604..817P}).  

We investigate the effect of embedding planets, brown dwarfs and low-mass main sequence stars into an envelope of a 3 $M_\odot$ star during various epochs of its evolution off of the main sequence.  That the secondary mass represents a small perturbation to the initial envelope configuration allows us to neglect detailed radiative and hydrodynamical effects.  We present as simplified a picture as possible in order to elucidate basic phenomonological consequences of the interaction.

\subsection{Envelope Binding Energy}
Our stellar model consists of a 3 $M_\odot$ main sequence star whose evolution is followed through the AGB phase with $X=0.74$ (mass fraction of hydrogen), $Y=0.24$ (mass fraction of helium), $Z=0.02$ and no mass loss (S. Kawaler - personal communication, Fig. \ref{machcs}).  A range of evolutionary models allows consideration of various positions and times at which the expanding envelope engulfs the orbiting brown dwarf.   We focus on three main epochs in the evolution: (i) near the tip of the first Red Giant Branch, (ii) the beginning of the Asymptotic Giant Branch and (iii) the quiet period between thermal pulses on the AGB branch.  In each case, we calculate the energy required to unbind the envelope mass above a given radius, $r$ (measured from the center of the primary's core) as follows:
\begin{equation}
E_{bind}(r)=-\int^{M_T}_{M}{\frac{GM(r)}{r}dm(r)},
\end{equation}
where $M_T$ is the total mass of the star and $M$ is the mass interior to the companions orbital radius.  Here it is assumed that the core and envelope do not exchange energy during the CE phase and that ejection of material has no bearing on core structure.  The values we determine for the binding energy in all three epochs are comparable to results from an estimation method first proposed by \cite{1984ApJ...277..355W} and further refined by \cite{2000A&A...360.1043D} and \cite{2001A&A...369..170T}.  Explicitly calculating the binding energy for each evolutionary epoch fixes our efficiency parameter $\alpha$ between 0 and 1.  

For the RGB star, our model core radius $r_c \sim3.5\times10^9$ cm with the envelope extending out to a radius $r_\star\sim7\times10^{11}$ cm.  At the chosen time in the RGB phase, the core contains 0.41 $M_\odot$ and the energy required to unbind the entire envelope ($\sim$ $10^{48}$ ergs) is the largest of our three epochs.  Once the star has ascended onto the AGB, the core contracts to a radius of 2.9 $\times$ $10^9$ cm and the envelope expands to $r_\star$ = 5.7 $\times$ $10^{12}$ cm.  The core has increased its mass to 0.55 $M_\odot$ and the energy required to unbind the entire envelope decreases to 1.3 $\times$ $10^{47}$ ergs.  For the interpulse AGB phase, the core has contracted to a $r_c\sim 1.6 \times 10^9$ cm while the envelope has expanded to $r_*\sim 1.3 \times 10^{13}$ cm.  The envelope binding energy has been further reduced to 5.7 $\times$ $ 10^{46}$ ergs with the core containing 0.58 $M_\odot$.  We find that the interpulse AGB phase is most favorable for binary induced envelope ejection since the range of masses and radii required to deposit favorable orbital energy into the envelope is greatest in this phase.  We discuss these results in detail in section 3.

\subsection{ Orbital Energy and Angular Momentum Evolution}
The immersion of the secondary in the envelope of the giant results in a reduction of the separation distance between core and companion.  To calculate the in-spiral and angular momentum transfer, we need to equate the rate of energy lost by drag to the change in gravitational potential energy.  The motion of a body under the influence of a central potential while incurring a drag force has been well studied and a general set of equations can be found in \cite{Pollard}.  Here we limit ourselves to the case where orbital eccentricity is negligible, such that the planet exhibits approximate Keplerian motion at each radii.  Under these conditions, the energy per unit time released by the secondary mass takes the following form:
\begin{equation}
L_{drag}=\xi\pi{R^2_a}\rho{(v-v_{env})}^3,
\end{equation}
where $v=\left(v_r,v_\phi,0\right)$ is the companion velocity, $\rho$ the envelope density, $v_{env}$ the envelope velocity, $R_a$ the accretion radius measured from the center
of the companion,  and $\xi$ is a dimensionless factor dependent upon the 
Mach number of the companions motion with respect to the envelope.  For supersonic motion, $\xi$ is greater than or equal to $2$ (\cite{1985MNRAS.217..367S}).  
The orbital motion of the planet is supersonic everywhere inside the orbit (see Fig. \ref{machcs}) and for simplicity, we assume a value of $\xi=4$.  The  value of $\xi$ acts only to slightly increase or decrease the in-spiral time of the secondary, while leaving the underlying physics unchanged.  The accretion radius is then given by
\begin{equation}
R_a\sim\frac{2Gm_2}{(v-v_{env})^2},
\end{equation} 
and represents the region around the secondary inside of which matter is gravitationally attracted to the secondary as it passes through the envelope.  If the companion moves close enough to the core, tidal effects can shred it.  We estimate the shredding radius (measured from the center of the primay's core) 
from balancing the differential gravitational force across the 
size of the companion $R_2$ (measured from the center of the companion) 
 with its self gravity, that is, 
 $\frac{d}{dr}\left({\frac{GM}{r^2}}\right)R_2\simeq\frac{Gm_2}{R_2^2}$, which yields $r_s\simeq\sqrt[3]{\frac{2M}{m_2}}R_2$.  

To determine the companion size, $R_2$, we separate our objects into three distinct groups: planets ($m_2\leq0.0026$ $M_\odot$; \cite{ZS}), brown dwarfs ($0.0026$ $M_\odot<m_2<0.077$ $M_\odot$; \cite{Burrows}) and low-mass main sequence stars.  We adopt an approximation to the models of \cite{Burrows}, identical to that used in \cite{RRL} for brown dwarfs, namely
\begin{equation}
R_2=\left[0.117-0.054{Log}^2\left(\frac{m_2}{0.0026}\right)+0.024Log^3\left(\frac{m_2}{0.0026}\right)\right]R_\odot.
\end{equation}
For low-mass main sequence stars, we adopt the homologous power-law used in \cite{RRL} given by
\begin{equation}
R_2=\left(\frac{m_2}{M_\odot}\right)^{0.92}R_\odot.
\end{equation}
As the separation between core and companion decreases, the secondary begins to fill its Roche lobe.  An approximation for the Roche lobe radius is given by (\cite{1983ApJ...268..368E}) 
\begin{equation}
R_{RL}=\frac{0.49q^{2/3}r}{0.6q^{2/3}+Ln(1+q^{1/3})},
\end{equation}
where $q$ is the ratio of secondary mass to core mass and $r$ is the binary separation distance.  Once $R_{RL}=R_2$, mass transfer ensues.

The time rate of change of gravitational potential energy of the binary is given by:
\begin{equation}
\frac{dU}{dt}=\frac{Gm_2v_r}{r}\left(\frac{dM}{dr}-\frac{M}{r}\right).
\end{equation}
As the secondary traverses the envelope, the drag luminosity must be supplied by the change in gravitational potential energy.  Therefore, we equate (3) and (8) using (4) and obtain an equation for the infall velocity.  This yields
\begin{equation}
v_r=\frac{4\xi\pi{G}m_2r\rho}{\left(\frac{dM}{dr}-\frac{M}{r}\right){\sqrt{v_r^2+\bar{v}_\phi^2}}},
\end{equation}
where $\bar{v}_\phi=v_\phi-v_{env}\simeq{v_\phi}$ for slowly rotating stars.  In addition, $v_r\ll{v_\phi}$ everywhere inside the envelope. Eq. (9) agrees with the limit of a general set of equations found in \cite{1976ApJ...204..879A} 
under these circumstances.
The time scale for infall from a position inside the envelope can then be estimated as $\tau\sim|\frac{r}{v_r}|$ (see Fig. \ref{infall}).  The in-spiral time is slightly shorter for the AGB star than for the interpulse AGB star.  In the outer reaches of the envelope, $\tau$ is comparable to the lifetime on the AGB ($\sim10^5$ yrs), but sharply drops to $\sim1$ yr just inside the outer radius.

As a consequence of the in-spiral process, the secondary loses orbital energy and angular momentum.  The reduction in orbital energy is given by:
\begin{equation}
\Delta{E_{orb}}(r)=\frac{GM_Tm_2}{2r_o}-\frac{GMm_2}{2r},
\end{equation}
where $r_o\sim{r}_\star$ is the stellar radius.  In practice, we expect $r_o$ to be slightly less then $r_\star$ since material in the outer reaches of the envelope exerts little  drag force, thereby significantly extending the infall time (see Fig. \ref{infall}).  In addition to transfer of orbital energy, as the secondary moves closer to the core, conservation of angular momentum results in a spin up of the initially stationary envelope.  We assume that the lost orbital angular momentum is transferred to spherical shells of the envelope.  In reality, most of the angular momentum may be confined close to the orbital plane resulting in latitudinal differential rotation in addition to that expected in the radial direction.  A more equatorially concentrated deposition of angular momentum could therefore lead to even more differential rotation
than considered herein, and a more robust dynamo.

The simple equations in this section  allow us to crudely investigate different outcomes of CE evolution.  Equations (1), (2), (5), (6) and (10) determine the depth at which various mass secondaries can penetrate into the star before depositing enough energy to unbind envelope material and Eq. (9) gives the radial component of velocity during in-spiral.  In the next section,
we discuss different CE end states that result from analyzing
these equations.

\section{Common Envelope Evolution Scenarios}
We outline three qualitatively different scenarios that can occur once the secondary is immersed in the envelope of a given stellar evolution phase: (i) the secondary provides enough orbital energy to directly unbind the envelope (or a portion of it) by itself, (ii) the secondary induces differential rotation in the envelope which can power a dynamo therein, unbinding the envelope or (iii) the secondary is shredded into a disc around the core, which can lead to a  disc driven outflow.  These scenarios are presented schematically  in Fig. \ref{threecases}.  Below we discuss each in depth and comment on their observational implications in Sec. 4.

\subsection{Secondary Induced Envelope Expulsion}
As the secondary enters the primary's envelope, the mutual drag transfers angular momentum to the latter and the secondary spirals in.  For an RGB star, envelope accretion onto brown dwarf secondaries was previously studied (\cite{1984MNRAS.208..763L}, \cite{1984MNRAS.210..189S}).  The stellar model consisted of a $0.88$ $M_\odot$ giant with a core mass of $0.72$ $M_\odot$ during hydrogen and helium shell burning phase.  The evolution of the giant was subsequently followed during in-spiral.  The authors found that the secondary evaporated if its mass was below an initial critical value ($m_{crit}\simeq0.005$ $M_\odot$).   When the
secondary exceeded this mass, it instead grew to $0.14$ $M_\odot$, independent of its initial supercritical mass.  The evolution was followed until the envelope was ejected, leaving a close binary system.  

For each evolutionary phase and fixed value of $\alpha$, we calculate the radius at which the orbital energy released equals that of the binding energy of the envelope for a range of secondary masses (see Fig. \ref{shred}).

 When the star has just reached the RGB phase, we find that no objects under $0.5$ $M_\odot$ can unbind the envelope before they are tidally shredded.  Even if the companion is not shredded, extracting enough orbital energy to expel the envelope requires penetrating to the inner most regions where physical contact with the core results.  Thus, we do not expect low-mass secondaries inside our model RGB star to produce helium white dwarfs.  

As the star enters its AGB phase, the envelope expands and the core contracts, thereby lowering the binding energy.  In this case, we find a range of masses and efficiencies for which the CE interaction can expel the envelope directly (without magnetic fields).  A $0.15$ $M_\odot$ companion provides enough orbital energy (for $\alpha=0.4$) to unbind the envelope when it reaches a radius of $3\times{10}^{10}$ cm.  For the interpulse AGB star, the binding energy is further reduced from the initial AGB, extending the range of masses and drag efficiencies that can unbind the system.  Even if only 20 percent of the orbital energy is available for envelope ejection ($\alpha =0.2)$, a $0.2$ $M_\odot$ secondary can expel the envelope at $5\times{10}^{10}$ cm.  The binding energy as a function of position is shown in Fig. \ref{Ebind}.  In addition, the orbital energy that a $0.02$ $M_\odot$ companion supplies as it traverses the envelope is also shown.

\subsection{Secondary Induced Envelope $\alpha-\Omega$ Dynamo}

Outflow production mechanisms that extract rotational energy may be required to explain the observed 
high power bipolar features of PN and PPNe, since radiation
driving is insufficient (Bujarrabal 2001).
Magnetic field generation inside the AGB envelope may provide a way of extracting this energy and collimating  an outflow.
Magnetically mediated outflows  in an isolated AGB star have been studied
from different perspectives (e.g. Pascoli 1993, \cite{EB2001}, \cite{2002MNRAS.329..204S}) and outflows from  binary + disc systems  
 (\cite{RRL}, \cite{1995MNRAS.273..146R}, \cite{1994ApJ...421..219S}, Blackman
et al. 2001b) have also been considered.  
Here we focus  on outflows from binary induced dynamos in 
the stellar envelope, and discuss accretion  driven outflows in the next section.

The AGB phase of stellar evolution provides the conditions needed to power an $\alpha-\Omega$ dynamo analagous to those studied in the sun, white dwarfs, and supernova progenitors (\cite{1993ApJ...408..707P},  \cite{1995ApJ...453..403T}, 
\cite{BNT}).  The combination of a deep convective envelope and differential rotation could generate large-scale magnetic fields.   \cite{EB2001} investigated an interface dynamo in the context of our $3.0$ $M_\odot$ AGB model.  They assumed that the star was initially rotating on the main sequence with a rotation rate of $200$ km/s.  Assuming that evolution off the main sequence conserves angular momentum on spherical shells  results in a differentially rotating AGB star (see Fig. \ref{rot}). Large-scale saturated fields of $B\sim5\times10^4$ G  
can then be calculated at the base of the convection zone.  
But to drive a magnetic outflow, the dynamo must operate over the entire lifetime of the AGB phase ($\sim10^5$ yrs) until enough matter has been radiatively 
bled from the envelope for the magnetic "spring" like a jack-in-the-box. 
 Unfortunately, field amplification drains energy from differential rotation and acts to transfer angular momentum from the core to the envelope, slowing down the core within 100 yrs.  Anisotropic convection 
does provide a possible mechanism for reseeding the differential rotation
(c.f. Rudiger 1989) and maintaining a steady
 AGB rotation profile (Blackman 2004), but 
more work is needed to assess the viability of the single star mechanism.  

	Alternatively, a binary companion may, via a CE phase, supply enough differential rotation to the envelope that the resulting amplified Poynting flux is large enough to unbind the envelope within a few years (Blackman 2004).  We study this concept more carefully here.  During the CE, the transfer of angular momentum and orbital energy to the envelope induces in-spiral of the companion.  Fig. 3 shows the envelope rotation profile produced when a $0.02$ $M_\odot$ brown dwarf travels from the outer part of the envelope to the core boundary at the beginning of the AGB phase.  The differential rotation profile from the single star approach of \cite{EB2001} is presented for comparison.  The magnitude of rotation generated from a binary interaction during the CE phase is a factor of 10 greater in the interface region and can therefore supply a significantly larger amount of differential rotation energy for a dynamo.  In principle, if the secondary could penetrate all the way to the core-envelope boundary (Fig. \ref{Ebind}), an additional region of energy could be tapped.  However, the penetration depth of the secondary is limited by tidal shredding, while the energy available for field amplification is constrained by how far the poloidal field can diffuse into the differential rotation zone (see \cite{BNT} for details).  

It should be noted that an interface dynamo will rapidly drain the available
differential rotation energy (Blackman, Nordhaus, Thomas 2006)
so unless the differential rotation is reseeded by convection,
the outflow produced by such a dynamo would be explosive and last 
$< 100$ years.  This is suggestive of ansae further comment on this in section 4.2.

For lower mass companions, the transfer of angular momentum may not produce strong enough differential rotation to drive a robust dynamo.  
The effect of a modest AGB dynamo (\cite{2001MNRAS.324..699S}) 
might be to  produce more sunspots near the stellar equator.  Dust formation is increased near these cool spots, enhancing the 
radiative mass loss rate there.  
If such a modest dynamo could last long enough, 
the formation of elliptical PNe might be aided by this mechanism. 

Further study of an $\alpha-\Omega$ interface dynamo produced from a secondary induced rotation profile is warranted (and in progress).  Preliminary results from our investigation of an interface dynamo operating in the AGB star model are encouraging.  For a differential rotation profile generated by the in-spiral of a $0.02$ $M_\odot$ companion (Fig. \ref{rot}), we obtain cycle periods of $\sim0.1$ yr with the transient dynamo lasting $0.5-1$ yr.

\subsection{Disc Driven Outflow}

In the event that the secondary cannot supply enough orbital energy to directly unbind the envelope or spin it up enough to power a dynamo, the secondary will be shredded from tidal forces as it nears the core.  The companion's physical radius, $R_2$, expands to fill its Roche lobe, at which point, mass transfer to the envelope ensues.  Eventually, near the core, tidal shredding occurs.  After several dynamical periods,  the remnant secondary mass forms an accretion disc which may be capable of producing collimated outflows (\cite{2001ApJ...546..288B}).  This scenario differs from \cite{1987PASP...99.1115M} in which the secondary strips material from the AGB primary and acquires a disc.  

\cite{1994ApJ...421..219S} investigated disc formation scenarios and found that a disc could form around the primary core when  a $\sim 1 M_\odot$ main sequence secondary is embedded in an AGB envelope.  At the end of the CE phase, after the primary has shed its envelope, the secondary expands, loses matter and forms a disc around the primary.  This is qualitatively similar to disc formation in cataclysmic variables, in which matter is stripped off a low-mass companion and forms a disc around a compact primary.  Such a disc may  be able to power collimated outflows during the proto-PNe phase.  This situation can occur if the companion directly ejected much of the envelope and avoided tidal shredding. 
In the present work, we focus only on low-mass companions, and 
on discs formed inside the CE from tidally destruction of
this companion.

\cite{RRL} investigated initial binary configurations which lead to  disc formation inside an AGB envelope from Roche lobe overflow of companions with masses between $0.001-1.0$ $M_\odot$.  Matter flowing through the inner Lagrangian point falls inward unless it has enough angular momentum to remain in Keplerian orbit.  For secondary masses above $0.05$ $M_\odot$,  matter stripped off the secondary falls all the way to the core surface and therefore does not form an accretion disc.  For massive planets and smaller brown dwarfs, an accretion disc can form.  

\cite{RRL} also study, the evolution of the resulting disc.  They  find that a geometrically thin accretion disc forms after an initial mass redistribution stage of
$\sim 1$yr.  For  a $0.03$ $M_\odot$ brown dwarf secondary with $\ge 10$\% 
 of its mass forming a disc, the accretion rate is found to be
 $\sim3\times10^{-7}\left[\left(\frac{t}{10^4}\right){^{-\frac{5}{4}}}\right]$ $M_\odot$ yr$^{-1}$.  For  a $3$ Jupiter mass secondary, the resulting disc is thinner and cooler with the accretion rate dropping to $4\times10^{-8}\left[\left(\frac{t}{10^4}\right){^{-\frac{5}{4}}}\right]$ $M_\odot$ yr$^{-1}$.  These estimates for mass flow  are comparable to those of young stellar objects where an accretion disc outflow connection has been established for similar values (\cite{1995ApJ...452..736H}).  

In Fig. 2, we show the distance from the core at which 
 various mass secondaries fill their Roche lobes or shred for our AGB and interpulse AGB model stars.  If the orbital energy deposited in the envelope is insufficient to unbid it, then the secondary continues migrating toward the core.  Tidal effects become increasingly important and we expect objects that penetrate deep enough to be tidally shredded into a disc (see Fig. \ref{shred}).  Brown dwarfs and massive planets shred to form a disc while low-mass stellar companions, because of their large radii ($\sim2\times10^{10}$ cm), may contact the core directly before forming a disc.

If the shredding of the companion results in an accretion disc (see Fig. \ref{threecases}), a disk outflow similar to those observed in other astrophysical objects such as young stellar objects, X-ray binaries and active galactic nuclei
is possible.

Disc driven outflows can sustain their
observable lifetime longer than the interface dynamos.
Thus extended bipolarities extending from the PPNe phase into the PNe phase
are suggested of disc mediated outflows rather than merely the explosive
outflow of an interface type dynamo discussed above.

\section{Discussion of Observational Implications}
As a consequence of our model, outflow composition, collimation and direction vary based on the mass of the secondary embedded in the envelope.  In section 4.1, we discuss observational implications of our three ejection scenarios (Fig. \ref{threecases}) and the possibility that they could  operate in conjunction.  In section 4.2, we comment on specific PPNe and PNe in the context of our CE framework.

\subsection{Observational Consequences}
The three basic ejection scenarios are shown in Fig. \ref{threecases}.  When the envelope is ejected purely via orbital energy deposition from the secondary, the corresponding PPNe outflows will reside primarily in the equatorial plane (see Fig. \ref{threecases}a).  There is numerical evidence from simulations that a binary induced equatorial outflow is confined to an opening angle of $\sim$ 20 - 30 degrees (\cite{1996ApJ...458..692T}).  \cite{1998ApJ...500..909S} follow the three-dimensional hydrodynamical evolution of an AGB star with companions of $0.4-0.6$ $M_\odot$.  The binary interaction funnels material and expels it along the equator.  If the rotation axis of the central star can be determined, then the identification of an equatorial outflow suggests a CE origin.  

Many PNe exhibit equatorial tori (\cite{2003AJ....126..848S}, \cite{1998ApJ...504..915B}, \cite{2002A&A...386..633C}), some of which are falling back towards the core.  This could be explained by a CE interaction that did not supply enough energy to unbind the envelope.  
Note that very small companions  fall into the core without
much envelope ejection. 
For very large mass companions, the radius at which the envelope is expelled increases, also resulting in a small amount of  matter in the equatorial outflow.
There is therefore an intermediate value of the companion mass which maximizes the amount of 
equatorial ejecta.

As compared to a high-mass, high density torus, a low-mass, low density torus may provide inadequate shielding from the central, illuminated white dwarf.  As a consequence, molecule formation in the equatorial torus is reduced.  
A companion that expels the entire envelope  would create more shielding
of the outer parts of the torus, leading to more molecule survival. 

In the event that the secondary cannot directly unbind material, the companion may induce differential rotation which ejects the envelope via a magnetic dynamo driven outflow (Fig. \ref{threecases}b). Such an outflow would be predominantly poloidal, likely collimated and aligned with the central rotation axis. The launching and shaping of the outflow could occur close to the central core and the role of a magnetic field at larger distances may be less important.  In addition, a torioidal  magnetic pressure ``sandwich'' across the equator acould squeeze some material out equatorially (Matt et al. 2004).  The overall outflow expected
from a magnetic outflow is thus predominantly poloidal with a smaller
equatorial component.

If the secondary is shredded into an accretion disc around the core, a disc driven outflow is possible (Fig. \ref{threecases}c).  In this case, the outflow may exhibit chemical signatures of the destroyed secondary.  The atmospheres of brown dwarfs are oxygen-rich, in contrast to certain carbon-rich AGB stars.  Water
and carbon monoxide 
 are present in a range of brown dwarf classes (\cite{2002ApJ...564..466G}, \cite{2002ApJ...564..421B}).  If the companion is a brown dwarf, a disc driven outflow can expel oxygen-rich material along the poles.  This may lead to the formation of crystalline silicates along the rotational axis.  If the secondary is a low-mass main sequence star, it may be difficult to detect any difference in outflow material if the primary is of similar composition.

As the CE phase evolves, a combination of the three above scenarios might occur.  For instance, differential rotation supplied during the CE phase may trigger a dynamo in the stellar envelope.  The companion continues its in-spiral and eventually forms a disc which later drives an outflow.  In this case, two winds are launched from the system, both along the polar axis.  The dynamo driven outflow is expected to occur in a burst, whereas the disc driven outflow might last $\simeq10^4$ yrs (\cite{2001ApJ...546..288B}).  Alternatively,  a companion may supply enough energy to directly unbind envelope material but subsequently shred into a disc.  The bulk of the mass is ejected along the equator while the disc material is ejected along the rotation axis.  

In short, each of the three possibilities in Fig. \ref{threecases} represent specific scenarios which may occur in conjunction or individually.  More work is needed to elucidate the detailed possibilities.

\subsection{Applications to specific PPNe and PNe systems}
\cite{Hsia} took time series photometric observations of the young planetary nebulae, Hubble 12 (HB 12; PN G111.8-02.8).  The authors found evidence for an eclipsing binary at the center in which the companion is a low-mass object ($m_2<0.443$ $M_\odot$).  In addition, there is evidence of reflection off of the secondary in the I and R bands.  The extended hourglass-like envelope of Hubble 12 suggests jet collimation.  In the context of our CE scenarios, such a collimated structure would result from either a dynamo driven outflow in the stellar envelope or a disc driven outflow around the progenitor core.  A low enough mass binary companion is required to produce a significant polar outflow.  
For $\alpha=0.6$ and a binary separation $r\sim8\times10^{10}$ cm, as suggested by \cite{Hsia}, 
the maximum mass that can result in a polar outflow without expelling much of the envelope equatorially  is approximately $ m_2\le 0.2M_\odot$.  For $\alpha=0.3$,
the limit is  $m_2\le 0.4$ $M_\odot$.   
If the AGB progenitor were more evolved when the CE phase commenced, then the upper limit on the corresponding masses would be lower.  For instance, if the companion were immersed in our interpulse AGB star, $\alpha=0.6$ would require $m_2\le0.08$ $M_\odot$ while $\alpha=0.3$ yields $m_2\le0.17$ $M_\odot$.  Therefore, it seems likely that the mass of the companion in Hubble 12 is at least a factor of $2$ or $3$ less then the upper limit proposed by \cite{Hsia}.

HD 44179, nicknamed the Red Rectangle, is a proto-PNe in which the secondary in the central binary is a low-mass post-AGB star (\cite{2002A&A...393..867M}).  The system consists of a disk with bipolar outflows emanating from the central region.  CO maps suggest that the circumstellar disk is in approximate Keplerian rotation (\cite{2003A&A...409..573B}).  In addition, the expanion velocity in the outer region is quite low ($\sim0.4$ km s$^{-1}$) suggesting that the disk is bound to the central binary.  It may be that equatorial matter ejected during a CE phase did not fully escape.  The disk material then fell backwards until the resulting angular momentum was sufficient to remain in a stable orbit.  

Recent Spitzer IRS data from the Red Rectangle show evidence of oxygen-rich material  in the carbon-rich bipolar outflows (\cite{2005ApJ...628L.119M}) in addition to the oxygen-rich material in the circumbinary disc (\cite{1998Natur.391..868W}).  
A possible evolutionary explanation for the disc composition, is that the progenitor incurred rapid equatorial mass loss while the star was still oxygen-rich.  The carbon-rich interior layers were exposed and used to shape the bipolar outflows.  The  oxygen-rich material in the carbon-rich outflows may be the result
of a jet from a disk composed of a shredded  brown dwarf or planet.  

NGC 7009 is a PNe, exhibiting complex morphology that includes two distinct ansae far from the central source.  \cite{2004ApJ...603..595F} analyzed the kinematics of the ansae and determined expansion velocities and proper motion.  For a distance of $0.86$ kpc to NGC 7009, the diameter of the ansae $\sim3.8\times10^{16}$ cm with a radial velocity of $\sim1.3\times10^7$ cm/s measured away from the nebulae.  This gives an upper limit for the burst time of the ansae ($\tau_a\simeq100$ yrs).  Thus an interface dynamo operating $\sim10^2$ years may be responsible for the production of ansae in some planetary nebulae.

\section{Conclusions}

We have studied the implications of 
embedding a range of low-mass $(m_2<0.3M_\odot)$ companions into the envelope of a $3.0$ $M_\odot$ star during three epochs of its evolution:  For RGB stars, we find that  envelope ejection is unlikely.  However, for  AGB and interpulse AGB stars, we find scenarios that can lead to partial or complete envelope 
ejection.  For an AGB star and conventional efficiency parameters ($0.3<\alpha<0.6$), we find that massive brown dwarfs can directly eject the envelope equatorially.  Lower mass companions  
 that do not directly eject the envelope  may spiral in 
far enough to induce a differential rotation mediated dynamo that 
ejects material poloidally.
In addition, the companion may be shredded into a disc, possibly facilitating a  disc driven outflow.  For the interpulse AGB star, the envelope has significantly expanded, further lowering the energy required to unbind the system.  In this phase, it is easiest for the envelope to be ejected.  


For systems in which the envelope is directly ejected, the expected outflow is equatorial with a torus-like appearance.  The amount of envelope material contained in the outflow is determined by the mass of the secondary and the penetration depth of the companion.  Shallow penetration depths may be indicative of higher mass companions and result in lower tori masses and less molecule formation rates in the expanding outflow.  Systems which form discs or incur dynamos are expected to generate polar outflows. 
If the companion is a brown dwarf that gets shredded inside the envelope of a carbon-rich AGB star, contamination of the polar outflow may result in the formation of crystalline silicates or other oxygen-rich substances.  

When the dynamo occurs in the envelope, via the induced
envelope differential rotation, and there is no reseeding of this differential rotation,  the outflow can only last  $< 100$yr. This would imply a poloidal
poweful but swift jet burst (e.g. ansae) in the PPNe phase. 
A disc dynamo may be   required for a disc mediated magnetic outflow 
but this would be powered by accretion, which falls off more gradually in time. 
The power from a disc mediated outflow
would therefore produce an observable outflow  over 
a longer time scale, and into the PNe phase (Blackman et al. 2001b).

To build on the current results, 
more detailed calculations are needed to include the  
 three dimensional nature of the binary interaction, the angular distribution of the induced outflow mass and composition, the operation of a dynamo, the inclusion of a wider range of companion masses, and the possibility of
 an initally rotating envelope.

\acknowledgements{We thank A. Frank, W. Forrest, J. Kastner, and 
I. Minchev for useful discussions and comments.  We would also like to thank S. Kawaler for use of his evolutionary models.
 JTN acknowledges financial support of a Horton Fellowship from the Laboratory for Laser Energetics through the Department of Energy and HST grant AR-10972.  
EGB acknowledges support from 
NSF grants AST-0406799, AST-0406823, and NASA grant ATP04-0000-0016
(NNG05GH61G).

{}
\begin{figure}
\plottwo{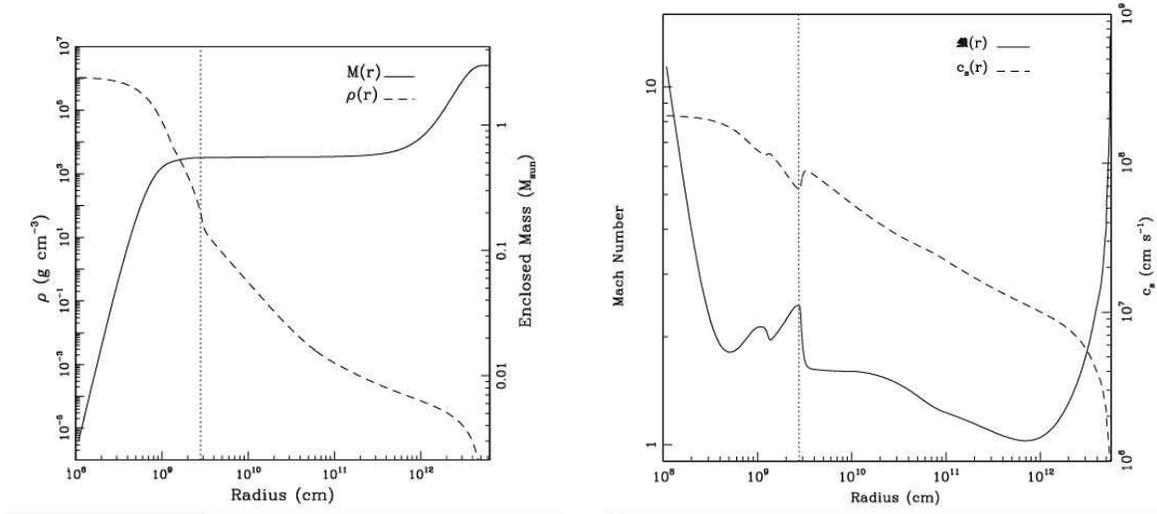}{MachSoundSpeed.epsf}
\caption{Left: Density and mass profiles for our model AGB star. The dotted line is the core-envelope boundary.  Right:  Mach number and sound speed as a function of radius.  The Mach number is computed from the Keplerian motion of the planet inside the envelope.  The motion is supersonic everywhere and thus justifies our choice of accretion radius (\cite{1952MNRAS.112..195B}).}
\label{machcs}
\end{figure}

\begin{figure}
\plottwo{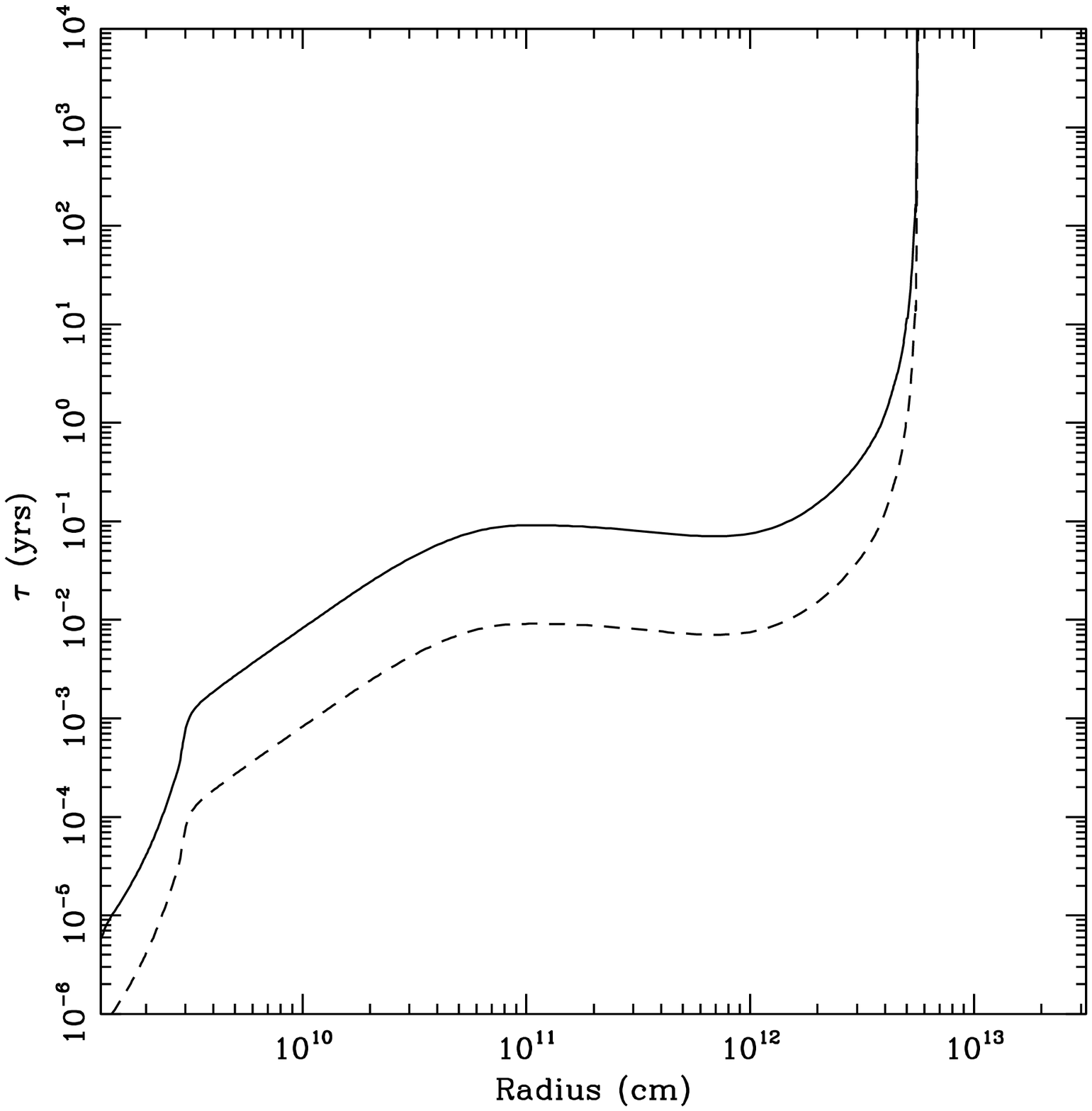}{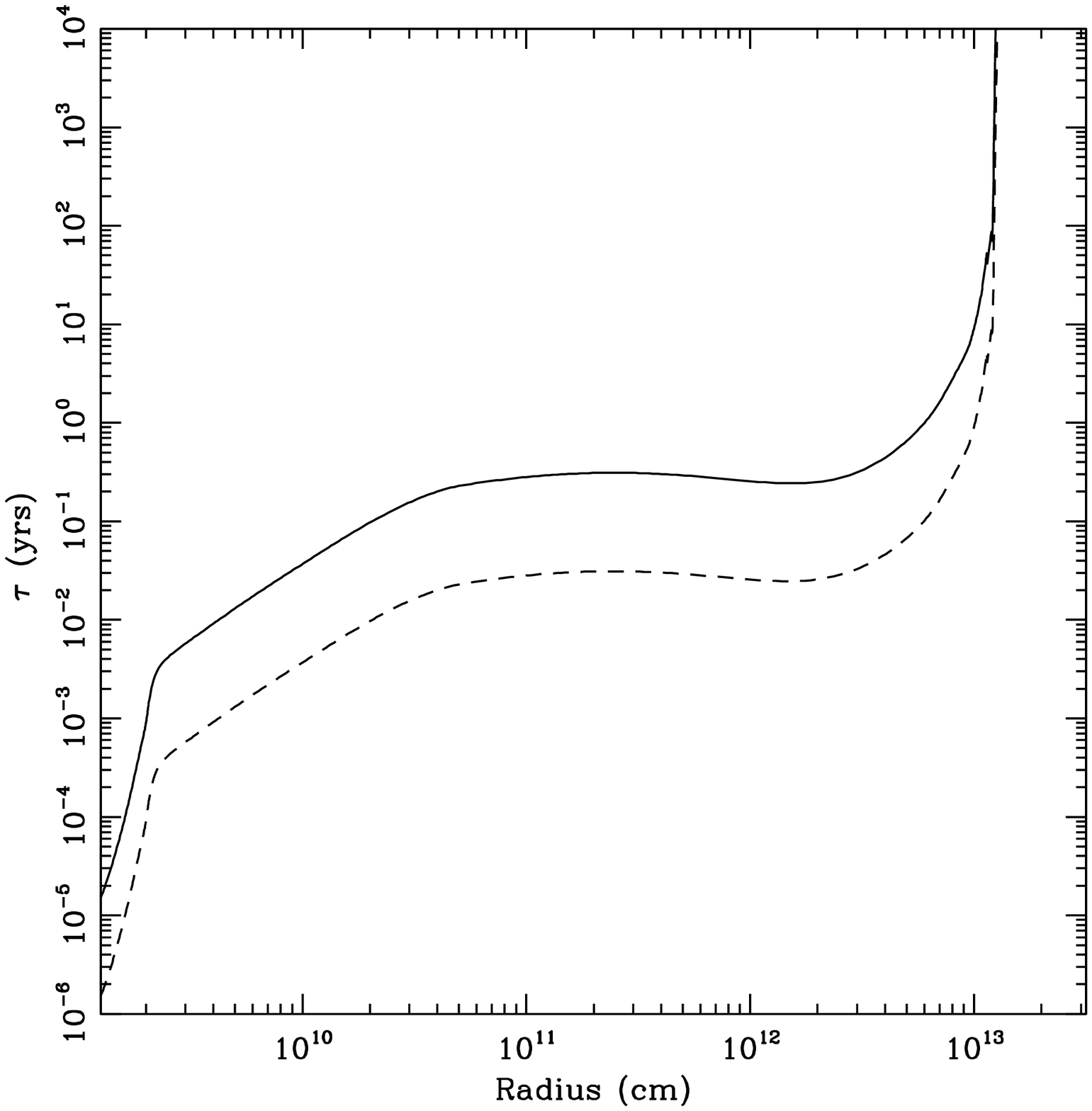}
\caption{Infall time as a function of position inside the envelope of the AGB star (left) and interpulse AGB star (right).  The solid line represents a companion of mass $0.02$ $M_\odot$ and the dotted line is a secondary of mass $0.2$ $M_\odot$.}
\label{infall}
\end{figure}

\begin{figure}
\epsscale{.7}
\plotone{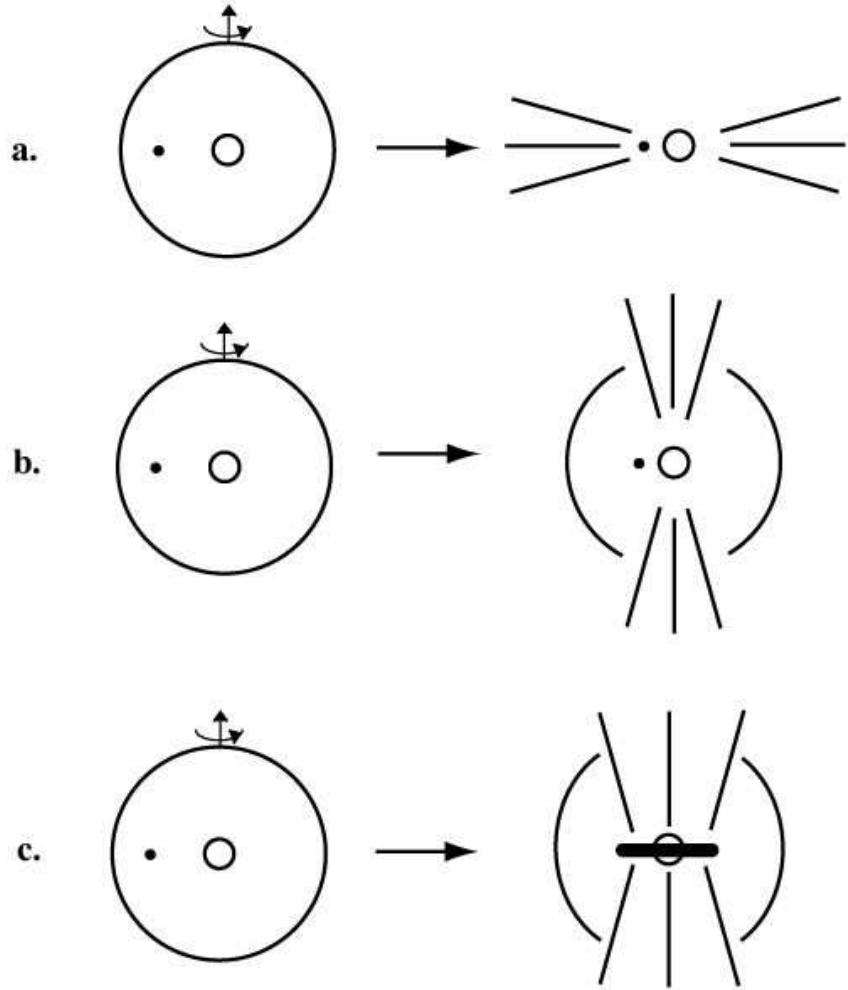}
\caption{Three possible outcomes of our CE evolution.  (a.) The companion becomes embedded in the stellar envelope, orbital separation is reduced, eventually resulting in unbinding the envelope equatorially.  (b.) The companion spirals in, the envelope is spun up causing it to differentially rotate.  The presence of a deep convective zone, coupled with the differential rotation, generates a dynamo in the envelope.  (c.)  The companion is shredded into an accretion disc around the core.  The disc then drives an outflow which, in principle, can unbind the envelope.}
\label{threecases}
\end{figure}

\begin{figure}
\plottwo{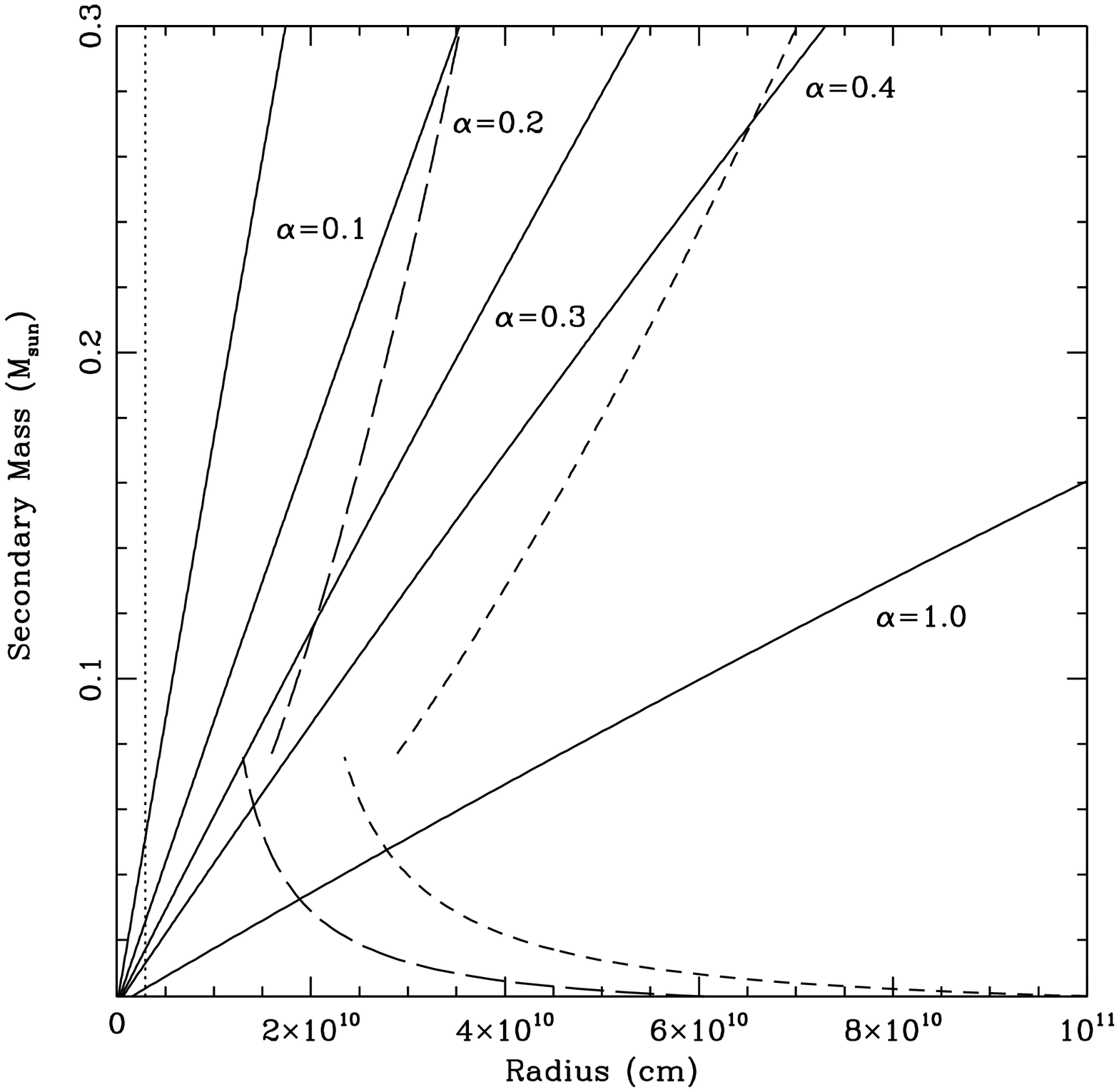}{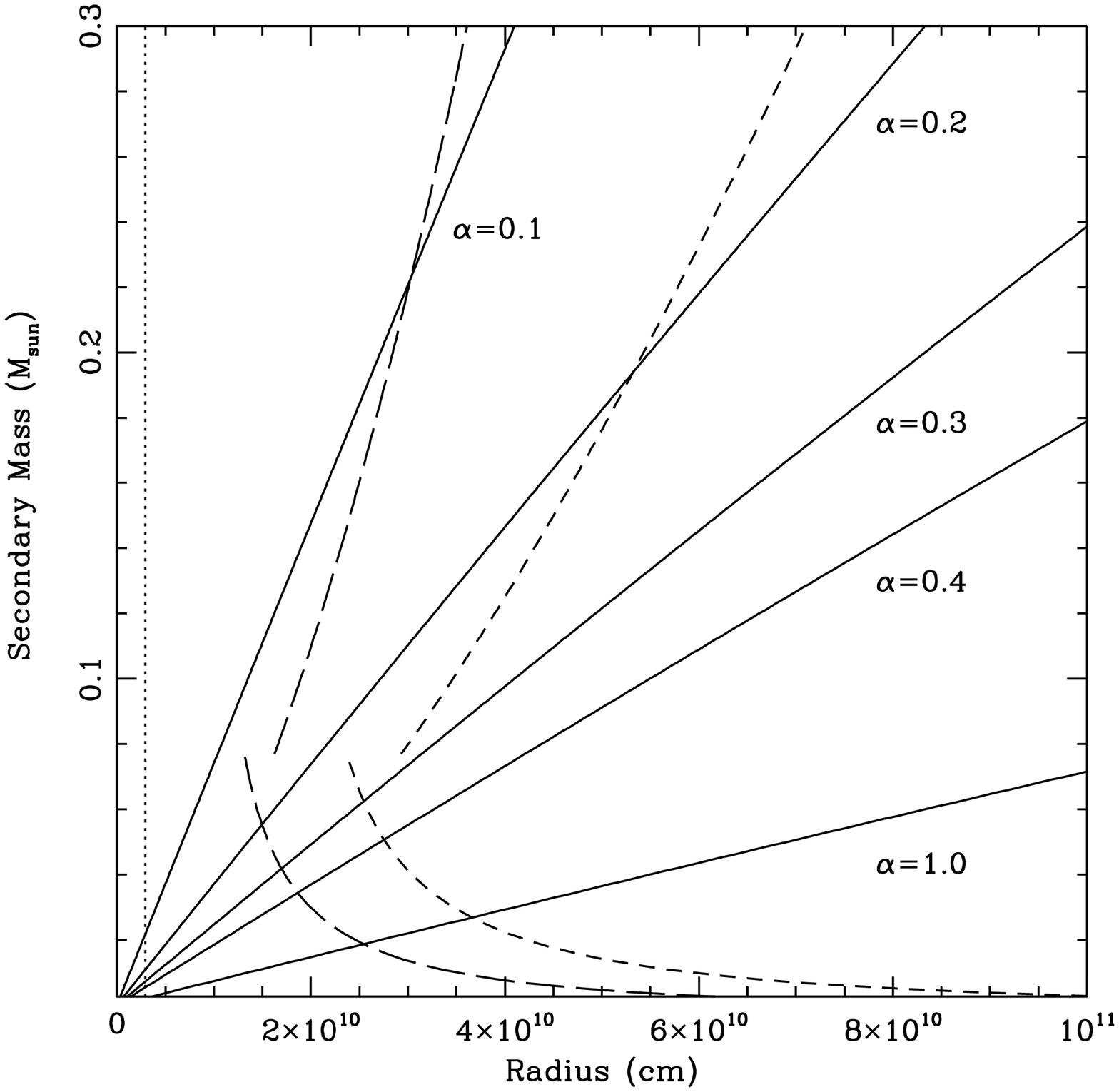}
\caption{For various efficiencies $\alpha$ (see Eq. 1), the solid line shows the radius at which the change in orbital energy equals the binding energy of the envelope for the beginning of the AGB star (left) and interpulse AGB star (right).  The dotted vertical line marks the core-envelope boundary.  The long-dashed line represents the radius at which the companion is tidally shredded by the core.  The short-dashed line is where the companion first fills its Roche lobe, initiating mass transfer to the envelope.}
\label{shred}
\end{figure}

\begin{figure}
\plottwo{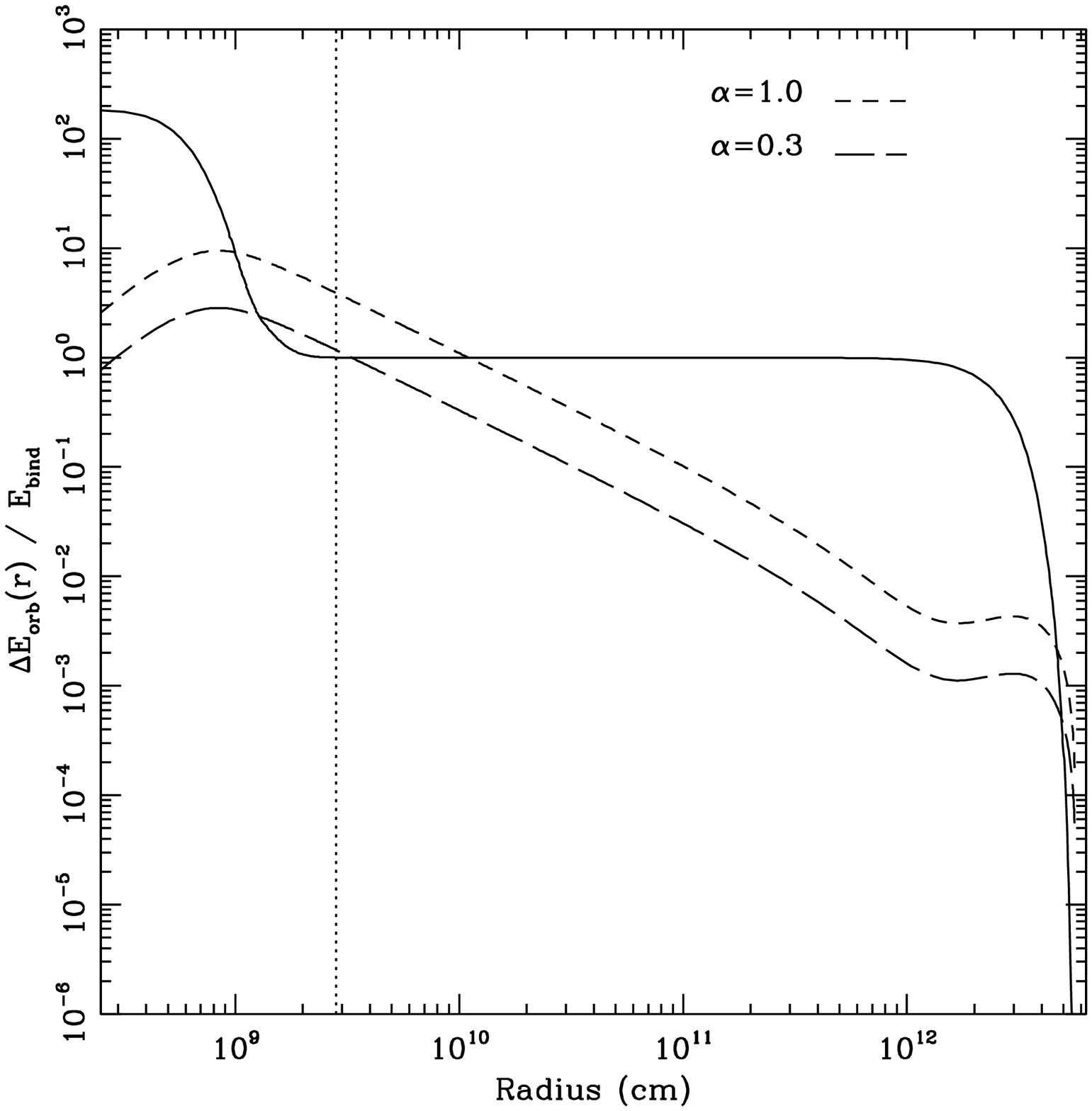}{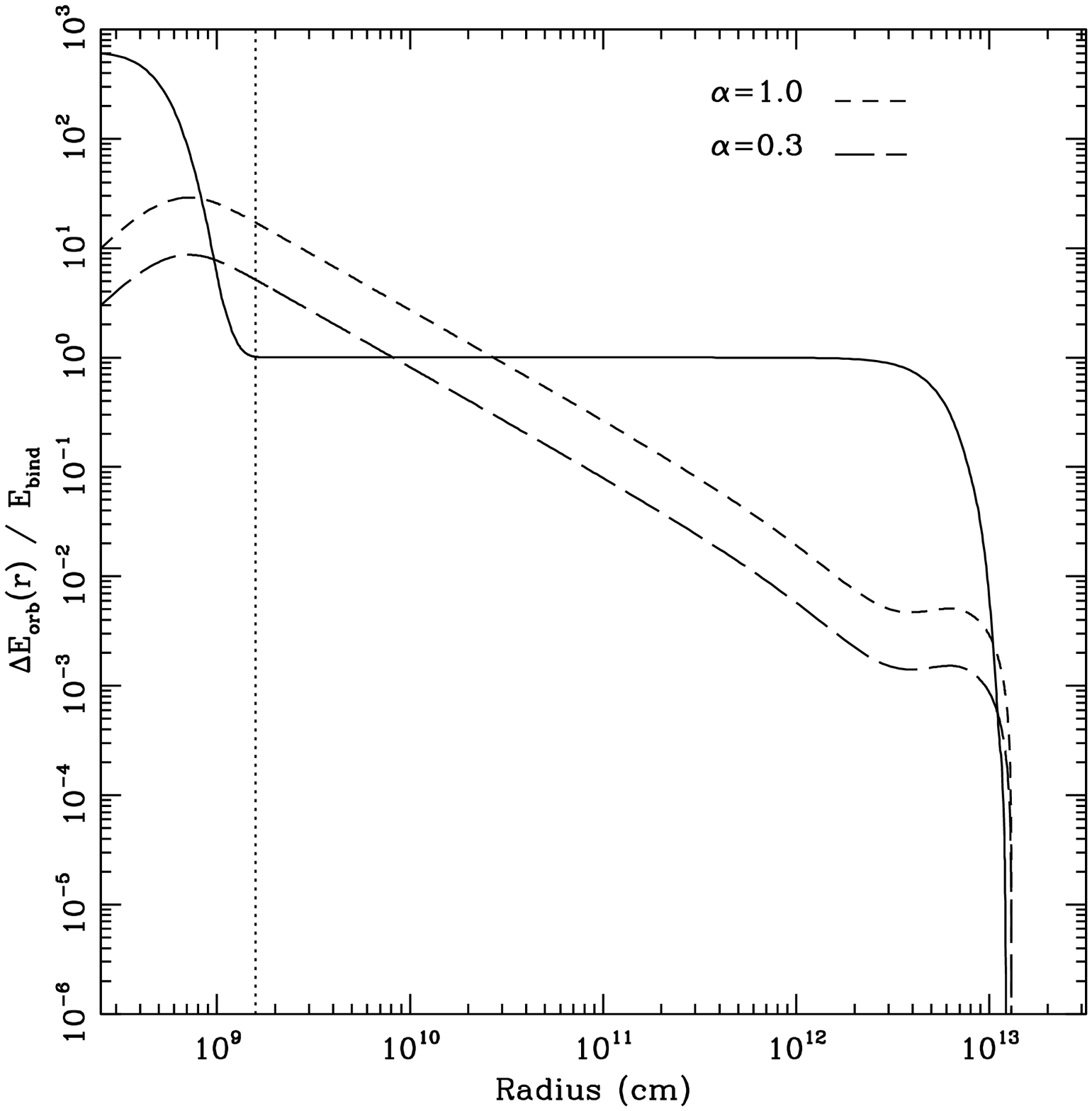}
\caption{The solid line depicts the energy required to unbind the envelope for the AGB star (left) and interpulse AGB star (right), if the secondary is not tidally shredded as it traverses the envelope.  The dashed lines represent the amount of energy deposited into the envelope from the change in orbital energy of the secondary for efficiency parameter $\alpha$ (Eq. 1).  For $\alpha=1.0$, a $m_2=0.02$ $M_\odot$ brown dwarf delivers enough energy to blow off the AGB  envelope at $r\sim10^{10}$ cm.  For $\alpha=0.3$, the brown dwarf must traverse all the way to the core-envelope boundary before supplying enough energy to unbind the system.  For smaller $\alpha$, a $m_2=0.02$ $M_\odot$ companion cannot unbind the AGB envelope before spiraling down to  a radius 
 where an interface dynamo might participate in unbinding the envelope.  For the interpulse AGB star, a 0.02 $M_\odot$ brown dwarf can supply enough orbital energy to unbind the envelope for $\alpha=1.0$ and $\alpha=0.3$.}
\label{Ebind}
\end{figure}

\begin{figure}
\plotone{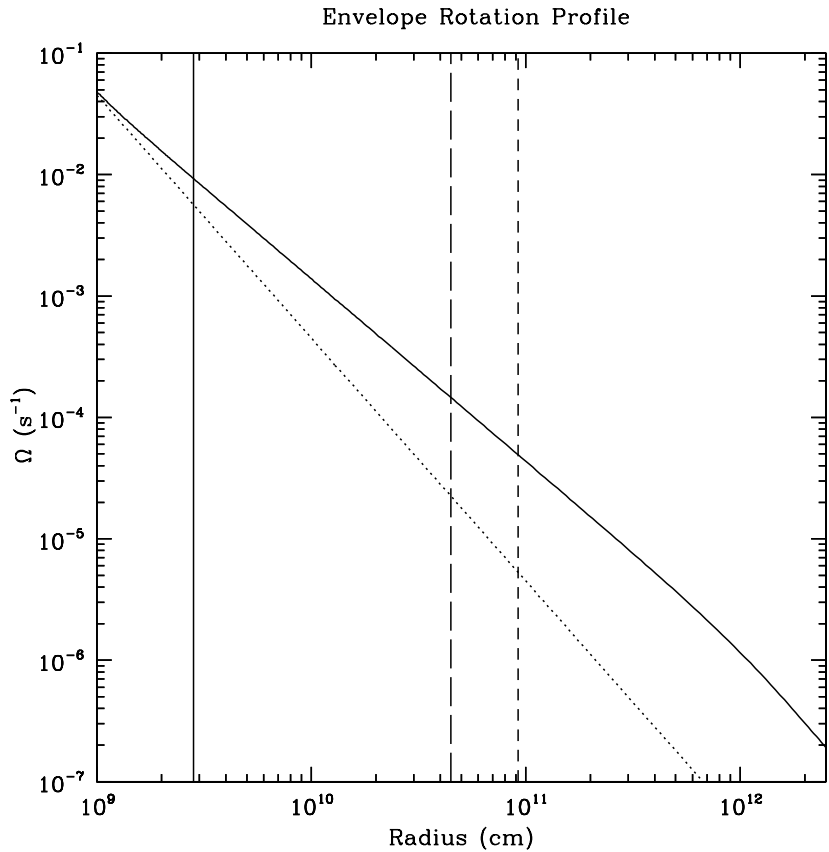}
\caption{Two rotation profiles for our 3.0 M$_\odot$ AGB star.  The solid curve represents the spin up of an initial stationary envelope by an infalling 0.02 M$_\odot$ brown dwarf.  The dotted curve is the rotation profile generated in \cite{EB2001} in which a main sequence star exhibiting solid body rotation conserves angular momentum of spherical mass shells during its evolution onto the AGB.  The solid vertical line marks the core boundary and the short-dashed line represents the base of the convective zone.  The long-dashed line is the base of the differential rotation zone used in \cite{EB2001}.}
\label{rot}
\end{figure}

\end{document}